\documentclass{ws-procs961x669}            
\begin{document}
\title{Neutron to Dark Matter Decay in Neutron Stars}

\author{T.F.Motta}

\address{CSSM and ARC Centre of Excellence for Particle Physics at the Terascale, Department of
	Physics, University of Adelaide SA 5005 Australia}

\author{P.A.M.Guichon}

\address{IRFU-CEA, Universit\'{e} Paris-Saclay, F91191 Gif sur Yvette, France}

\author{A. W. Thomas$^*$}

\address{CSSM and ARC Centre of Excellence for Particle Physics at the Terascale, Department of
	Physics, University of Adelaide SA 5005 Australia}

\begin{abstract}
Recent proposals have suggested that a previously unknown decay mode of the neutron into a dark matter particle could solve the long lasting measurement problem of the neutron decay width. We show that, if the dark particle in neutron decay is the major component of the dark matter in the universe, this proposal is in disagreement with modern astro-physical data concerning neutron star masses. 
\end{abstract}


\bodymatter

\section{Introduction}
In a recent publication by Fornal \textit{et al.}~\cite{Fornal} a proposal was made to solve the persistent discrepancy between two methods of measuring the neutron life-time. Trapped neutrons in a bottle appear to have a shorter life-time than neutrons in a beam where the decay proton is detected. There is a discrepancy of around 8 seconds (3.5$\sigma$) between the two experimental set-ups. In Ref.\cite{Fornal} it was suggested that the reason for this difference could lie in a formerly unknown decay channel of the neutron to a dark fermion. This proposal came as an alternative to the previous hypothesis that the experimental disagreement could be caused by the neutron oscillating into it's mirror counterpart \cite{Mirror}. Both arguments rely on the proposed existence of a decay channel to a fermion almost degenerate with the neutron.

This proposal attracted the attention of several collaborations and a number of publications followed the original release. In Ref.\cite{Tang} the authors argue through experimental evidence that in a decay of the form $n \rightarrow \text{DM} + \lambda$, i.e. a dark matter particle plus another decay product $\lambda$, that extra particle could not be a photon ($\lambda \not= \gamma$). Another publication\cite{Antineutrino} pointed out that this hypothesised decay could also explain a different experimental inconsistency, the ``reactor antineutrino anomaly", that is, the $3\sigma$ discrepancy between theory and measurement of the antineutrino flux from a reactor.
Finally Czarnecki \textit{et al.}~\cite{Czarnecki} although they did not rule out this explanation, pointed out strong constraints related to the value of the neutron axial charge.

In this publication we argue that allowing the neutron to decay to an almost degenerate dark fermion would mean that inside a neutron star, where the neutrons occupying a Fermi sea can sustain, through degeneracy, very large pressures, a large portion of these neutrons would decay to this dark fermion. This implies a severe decrease in pressure, which means that the maximum mass of neutron stars before gravitational collapse would be drastically lower than the masses of the stars measured so far. This was argued in Refs.~\cite{Motta:2018rxp,Baym,Reddy} and will be developed in further detail in this publication.

\section{Framework}

Simulating the internal structure of neutron stars ultimately amounts to solving the so-called Tolman-Oppenheimer-Volkof\cite{TOV} (TOV) equations for several different values of central energy density. The TOV equations give an internal profile for the pressure of the star through ($c=G=\hbar=1$)
\begin{eqnarray}\label{tov}
\frac{dP(r)}{dr}=-\frac{1}{r^2}\left ( \epsilon(r) + P(r) \right )\left ( M(r)+4\pi r^3 P(r) \right )\left ( 1-\frac{2M(r)}{r} \right )^{-1}
\end{eqnarray}
and the mass is given by the continuity equation
\begin{equation}
\frac{dM(r)}{dr}=4\pi r^2\epsilon(r).
\label{massagain}
\end{equation}
This set of equations take, as an input, the equation of state (EOS) of the matter of which the star is made.
We will adopt, as a model for the core of neutron stars, the infinite nuclear matter EOS from the quark-meson coupling model\cite{QMC} recently reviewed in Ref.\cite{QMC_Review}. This model is well established and has been shown to provide an adequate description of high density nuclear matter in several previous calculations\cite{QMC1,QMC2}.

We compare that equation of state with a modified version of it where the neutron decays to a dark fermion. Since a difference in mass of the order of a few MeV makes absolutely no difference to the mass of a neutron star, we will take this dark fermion to be fully degenerate with the neutron. Ultimately we will show that adding a vector self interaction among the dark fermions can indeed bring the mass up to more acceptable values, as was also shown in Ref.~\cite{Vector}. However, in order for that to happen, the coupling of this vector intermediate particle with the dark fermion has to be simply huge and we will argue that recent publications~\cite{Barbecue,DAmico} rule out that explanation if the dark particle in neutron decay is the major component of the dark matter in the universe.

\subsection{Dark Matter}

The proposal by Fornal \textit{et al.}~\cite{Fornal} is based on the decay of the neutron into a dark matter fermion which is almost degenerate with the neutron itself, plus another lighter component to conserve energy.

Their first of three proposals mentioned in the publication is $n \rightarrow \chi + \gamma$, where $\chi$ is (and hereafter refers to) the dark matter fermion. However, as argued above, this model was experimentally excluded by Tang \textit{et al.}~\cite{Tang}. The only viable mode seems to be
\begin{eqnarray}
	n\rightarrow \chi + \phi
\end{eqnarray}
where $\phi$ is a much lighter dark boson. This requires that the energy of the dark particles be in the ranges
\begin{eqnarray}
&937.900\text{MeV}<m_\chi<938.543\text{MeV}\\
&937.900\text{MeV}<m_\chi+m_\phi<939.565\text{MeV}.
\end{eqnarray}

We argue that
\begin{romanlist}
	\item In neutron stars, the presence of this light dark boson $\phi$ is completely irrelevant for it would escape the system very quickly.
	\item All of the proposed models indicate that, in neutron stars, the only change this hypothesis implies is a change in chemical composition from the equilibrium reaction $n \leftrightarrow \chi$, here imposed by the chemical equilibrium equation for the chemical potentials $\mu_n=\mu_\chi$.
\end{romanlist}

\subsection{QMC}
The chosen model of nuclear matter interaction is the QMC model\cite{QMC}. Based on a quark description of the baryons as quark bags interacting directly with mesons (scalar-isoscalar $\sigma$, vector-isoscalar $\omega$, vector-isovector $\rho$) we derive the energy density of the system in Hartree-Fock (HF) approximation. 

The Hartree, or mean field, contribution amounts to
\begin{align}
	&\epsilon_\text{Hartree}=\frac{m_\sigma^2\sigma^2}{2} + \frac{m_\omega^2\omega^2}{2}+ \frac{m_b^2b^2}{2}& \nonumber\\
	&+\frac{1}{\pi^2}\int_{0}^{k_F^n}{k^2}{\sqrt{k^2+M_N^*(\sigma)^2}dk} +
	\frac{1}{\pi^2}\int_{0}^{k_F^p}{k^2}{\sqrt{k^2+M_N^*(\sigma)^2}dk}  \nonumber\\
	&+\frac{1}{\pi^2}\int_{0}^{k_F^e}{k^2}{\sqrt{k^2+m_e^2}dk} +
	\frac{1}{\pi^2}\int_{0}^{k_F^\mu}{k^2}{\sqrt{k^2+m_\mu^2}dk}
	+\frac{1}{\pi^2}\int_{0}^{k_F^\chi}{k^2}{\sqrt{k^2+m_\chi^2}dk}
\end{align}
where the effective mass of the nucleon is $M_N^*(\sigma)=m_n-g_\sigma\sigma+\frac{d}{2}(g_\sigma\sigma)^2$. The $d$ is what is refered to as scalar polarizability and it is a prominent feature of the QMC model. In our convention $\sigma$, $\omega$, and $b$ refer to the mean field values of the mesons (where $b$ is the mean field value of $\rho$). For each particle the fermi momenta and chemical potentials as functions of the number densities are calculated as
\begin{align}
	&k_\varphi^{3}={{3\pi^2n_\varphi}},\quad \varphi=\{p,n,e,\mu,\chi \} \\
	&\mu_n= \frac{\partial \epsilon}{\partial n_n}, \quad \mu_p=  \frac{\partial \epsilon}{\partial n_p}, \quad \mu_l= \sqrt{k_f(n_l)^2+m_l^2}
\end{align}
And finally the Fock terms 
\scriptsize\begin{align}
&\epsilon_\text{Fock}=-G_\omega\frac{1}{(2\pi)^6} \left[
\int_0^{k_F^p}d^3k_1 \int_0^{k_F^p}d^3k_2 \frac{m_\omega^2}{(\vec k_1 - \vec k_2)^2 + m_\omega^2} 
+
\int_0^{k_F^n}d^3k_1 \int_0^{k_F^n}d^3k_2 \frac{m_\omega^2}{(\vec k_1 - \vec k_2)^2 + m_\omega^2}\right] \nonumber\\
&-\frac{G_\rho}{4}\frac{1}{(2\pi)^6} \left[
(1)\times\int_0^{k_F^n}d^3k_1 \int_0^{k_F^n}d^3k_2 \frac{m_\rho^2}{(\vec k_1 - \vec k_2)^2 + m_\rho^2} +(1)\times\int_0^{k_F^p}d^3k_1 \int_0^{k_F^p}d^3k_2 \frac{m_\rho^2}{(\vec k_1 - \vec k_2)^2 + m_\rho^2}\right. \nonumber\\
&\left. 
+ (2)\times\int_0^{k_F^n}d^3k_1 \int_0^{k_F^p}d^3k_2 \frac{m_\rho^2}{(\vec k_1 - \vec k_2)^2 + m_\rho^2}+ (2)\times\int_0^{k_F^p}d^3k_1 \int_0^{k_F^n}d^3k_2 \frac{m_\rho^2}{(\vec k_1 - \vec k_2)^2 + m_\rho^2} \right]&\nonumber\\
& +\frac{1}{(2\pi)^6}
\int_0^{k_F^p}d^3k_1 \int_0^{k_F^p}d^3k_2 \frac{1}{(\vec k_1 - \vec k_2)^2 + \tilde m_\sigma^2}\times\frac{M_N^*(\sigma)(-g_\sigma C(\sigma))}{\sqrt{M_N^*(\sigma)^2+k_1^2}}
\times\frac{M_N^*(\sigma)(-g_\sigma C(\sigma))}{\sqrt{M_N^*(\sigma)^2+k_2^2}}\nonumber\\
&+\frac{1}{(2\pi)^6}
\int_0^{k_F^n}d^3k_1 \int_0^{k_F^n}d^3k_2 \frac{1}{(\vec k_1 - \vec k_2)^2 + \tilde m_\sigma^2}\times\frac{M_N^*(\sigma)(-g_\sigma C(\sigma))}{\sqrt{M_N^*(\sigma)^2+k_1^2}}
\times\frac{M_N^*(\sigma)(-g_\sigma C(\sigma))}{\sqrt{M_N^*(\sigma)^2+k_2^2}}\nonumber
\end{align}\normalsize
where
\begin{align}
	\tilde m_\sigma^2 =m_\sigma^2 + \frac{1}{\pi^2}\sum_{p,n}\int_0^{k_f^n}k^2dk \frac{\partial^2}{\partial \sigma^2} \sqrt{M_N^*(\sigma)^2+k^2}.
\end{align}

The density dependent meson mean field equations in the QMC model are
\begin{align}
&\sigma(n_n,n_p) =- \frac{1}{m_\sigma^2\pi^2}\left( \frac{\partial M_N^*}{\partial\bar\sigma}\right)\left[\sum_{p,n}
\int_{0}^{k_F}k^2dk\frac{M_N^*(\sigma)}{\sqrt{k^2+M_N^*(\sigma)^2}}
\right], \\
&\omega(n_n,n_p)=\frac{g_\omega}{m_\omega^2}\left(n_n+n_p\right), \\
&b(n_n,n_p)=\frac{g_\rho}{m_\rho^2}\left(\frac{n_p}{2} -\frac{n_n}{2}\right).
\end{align}
and finally the pressure is calculated as $P=\sum_f\mu_fn_f-\epsilon$.

In Table~\ref{tab:constants} we report the constants used to perform the calculations. They are chosen to fit the saturation density at $0.16\text{fm}^{-3}$, the binding energy of symmetric matter at saturation $-15.8\text{MeV}$ and symmetry energy $30\text{MeV}$.

\begin{table}
	\tbl{Masses and coupling constants.}
	{\begin{tabular}{@{}lcccccccc@{}}
			\toprule
			\ &$\sigma$ & $\omega$ & $\rho$ & $n$ & $p$ & $e$ & $\mu$ & $\chi$ \\
			Mass &700MeV&782MeV&775MeV&939MeV&939MeV&0.5MeV&105MeV&939MeV \\
			Coupling ($g^2/m^2$) &11.33fm$^2$&7.27fm$^2$&4.56fm$^2$&.&.&.&.&.
		\end{tabular}
	}
	\label{tab:constants}
\end{table}

\section{Neutron Stars}

Using the model presented above we calculate the equilibrium densities through the equations
\begin{align}
	\text{Neutron $\beta$ decay}\quad&\mu_n=\mu_p+\mu_e\\
	\text{Muon $\beta$ decay}\quad&\mu_\mu=\mu_e \\
	\text{Charge neutrality}\quad&n_p = n_e+n_\mu \\
	\text{Dark matter decay}\quad&\mu_n = \mu_\chi.
\end{align}

\begin{figure}[ht]
\begin{center}
	\includegraphics[width=4in]{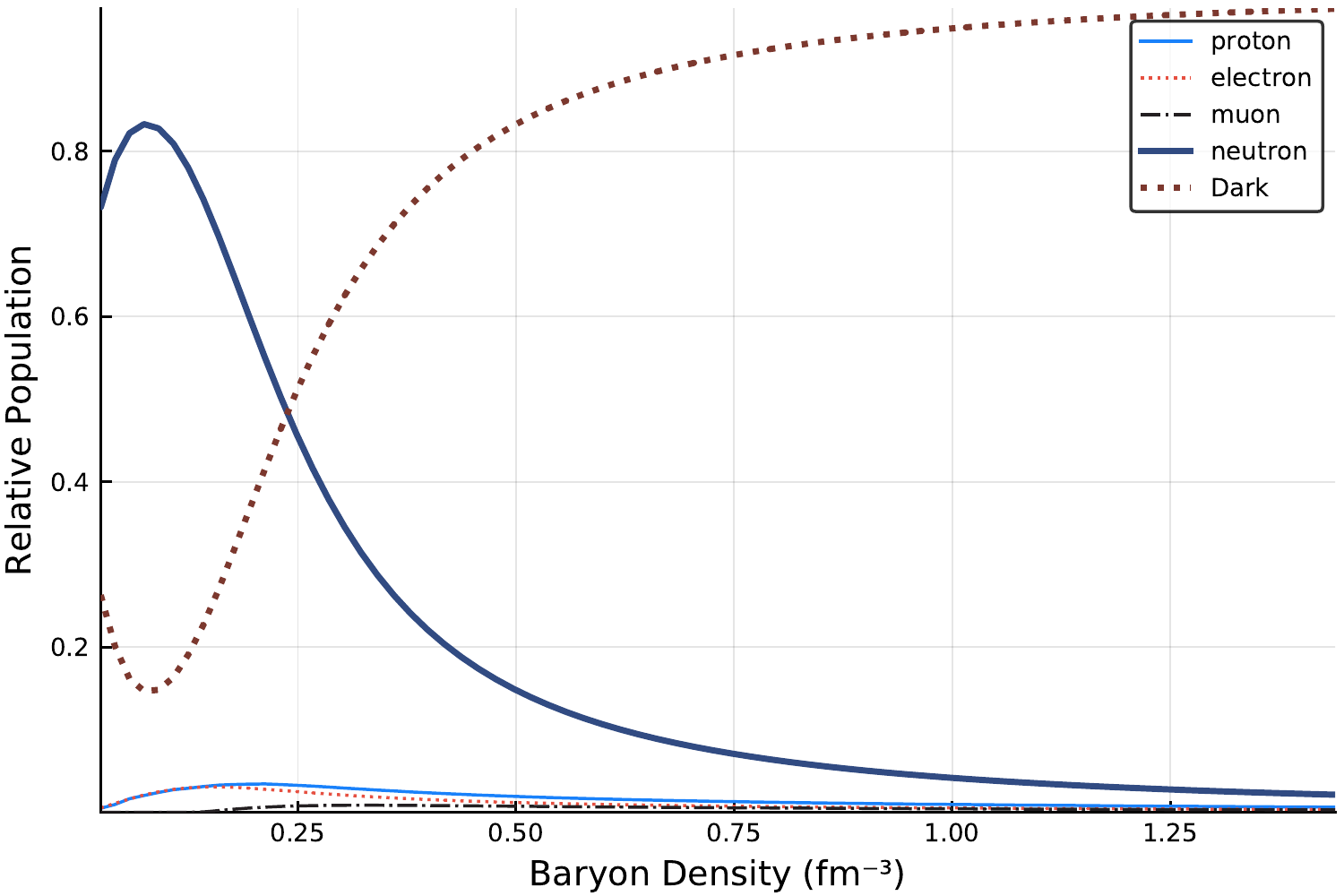}
	\caption{Relative species fraction with dark matter present.}
	\label{fig:fraction}
\end{center}
\end{figure}

\begin{figure}[ht]
	\begin{center}
	\includegraphics[width=4in]{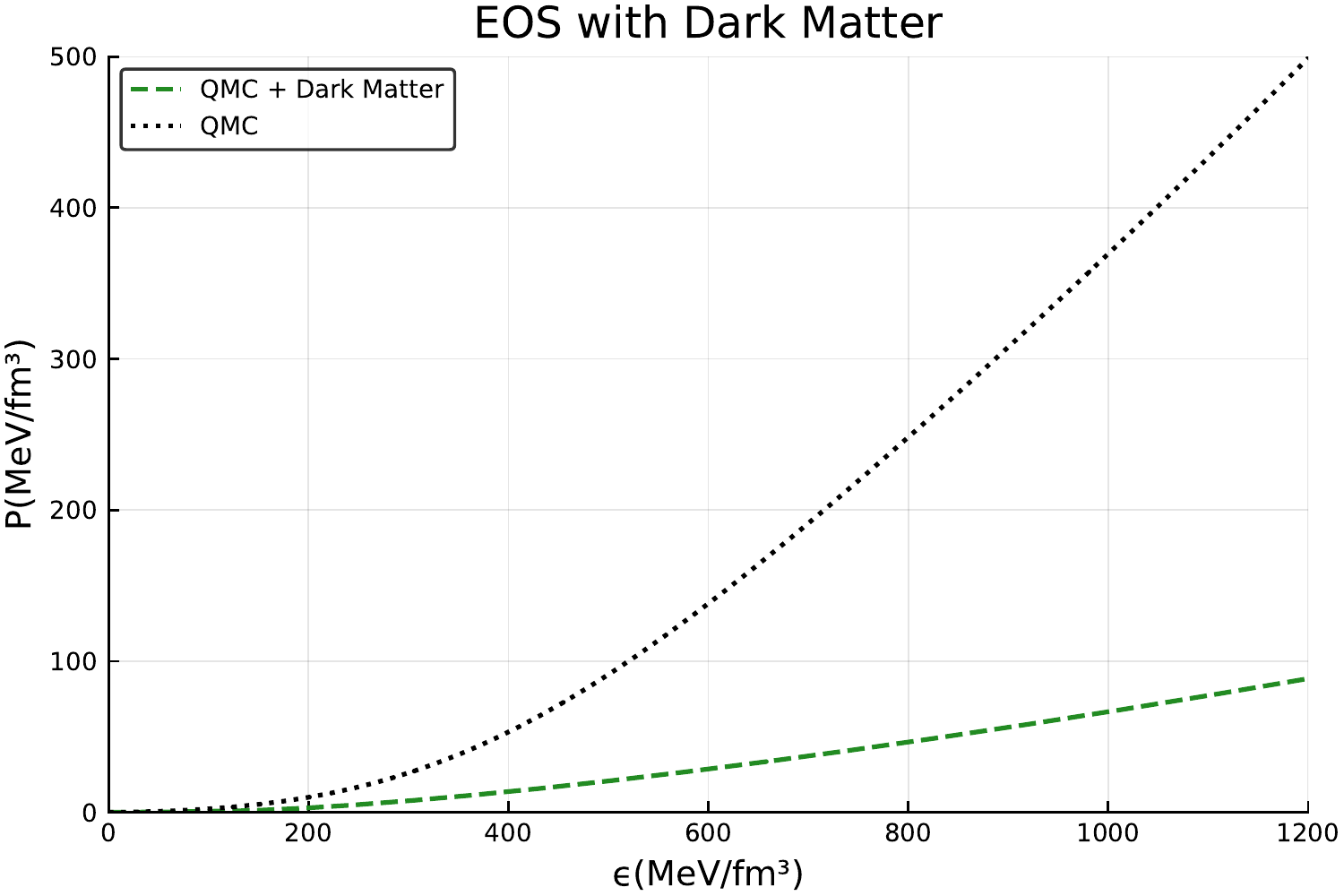}
	\caption{Equation of state with dark matter present.}
	\label{fig:eos}
	\end{center}
\end{figure}

\begin{figure}[ht]
	\begin{center}
		\includegraphics[width=4in]{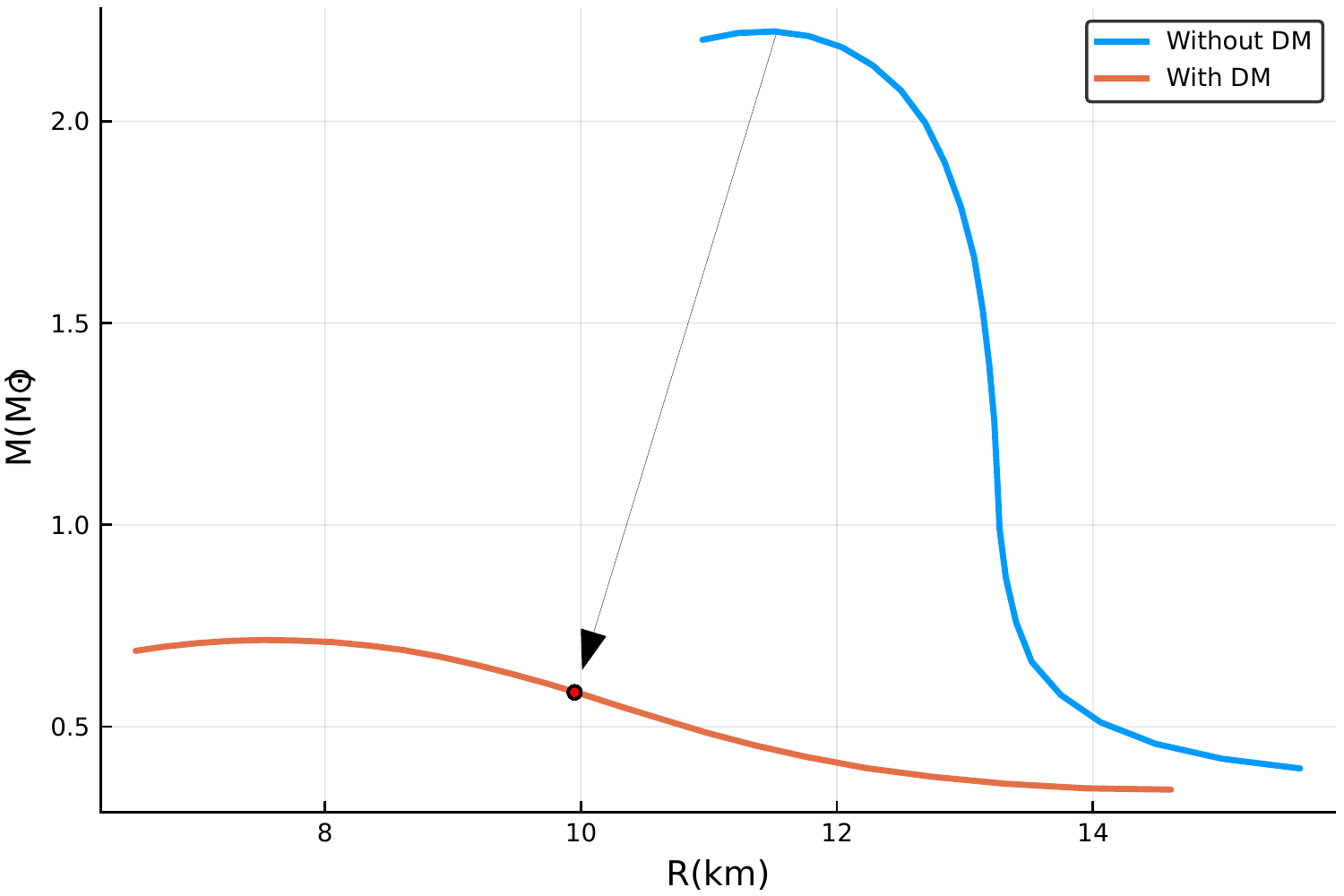}
		\caption{Maximum possible star as an end product of dark matter decay.}
		\label{fig:massaraio}
	\end{center}
\end{figure}

Solving these equations we get species fractions that vastly favours the dark matter particle 
(Fig.~\ref{fig:fraction}).
That, in turn, leads to a drastic decrease in pressure in the equation of state (Fig.\ref{fig:eos}) and as a consequence the Mass versus Radius diagram has a maximum significantly lower than the case without dark matter (Fig.\ref{fig:massaraio}).

\begin{figure}[ht]
	\begin{center}
		\includegraphics[width=2.5in]{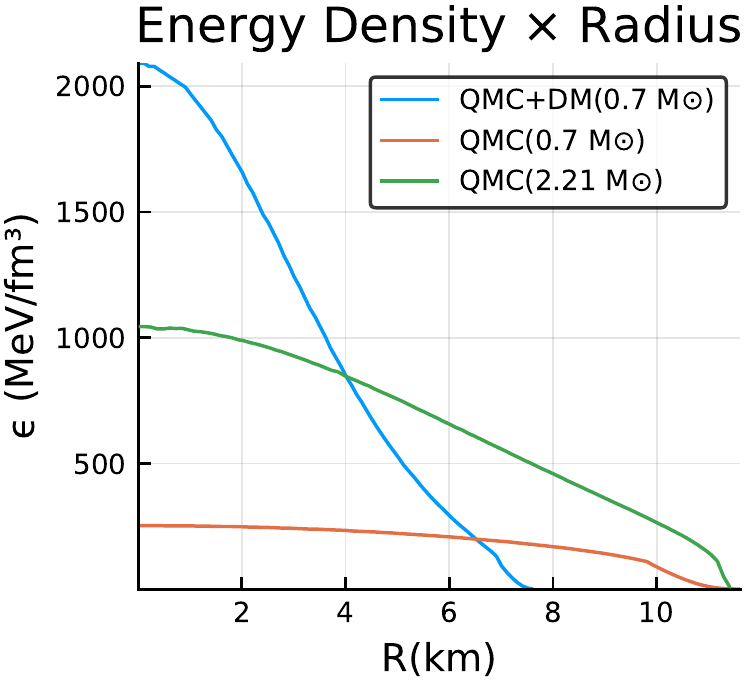}
		\caption{Energy density profile as a function of the internal radius of the star. Profile is presented for the maximum mass point of the diagram with dark matter present, an equally heavy star without dark matter and the maximum mass star without dark matter.}
		\label{fig:edens}
	\end{center}
\end{figure}

The maximum mass in the diagram with dark matter is of around $0.7$M$_\odot$. The reason for that is that although the central energy density of a star with dark matter is much larger than the star without it, it does not have enough pressure to support itself and therefore the energy density goes down very quickly (Fig.\ref{fig:edens})
However, it is unreasonable to assume that even the upper most point in the mass radius diagram of the EOS with dark matter present could ever be reached. Since the star with dark matter has to come from a real star we take the maximum mass star without dark matter and check to which point in the diagram the decay star would occur, that is, which point in the dark matter diagram has a total baryon number plus total dark matter number equal the total baryon number of a star without dark matter. That leads to a maximum mass of $0.58$M$_\odot$ (Fig.\ref{fig:massaraio})

\section{Repulsive Vector Interaction}
If the dark matter particle were self interacting through a repulsive interaction it is possible that it could build up pressure to sustain larger masses. This approach was used in Ref.\cite{Vector} and we here perform the same procedure within the framework of QMC. To compare with the neutron-$\omega$ physical system we vary the coupling/mass as multiples of the $n\omega$ vertex couplings,  as indicated in the figures. We name this vector intermediate $V$.

\begin{figure}[ht]
	\begin{center}
		\includegraphics[width=4in]{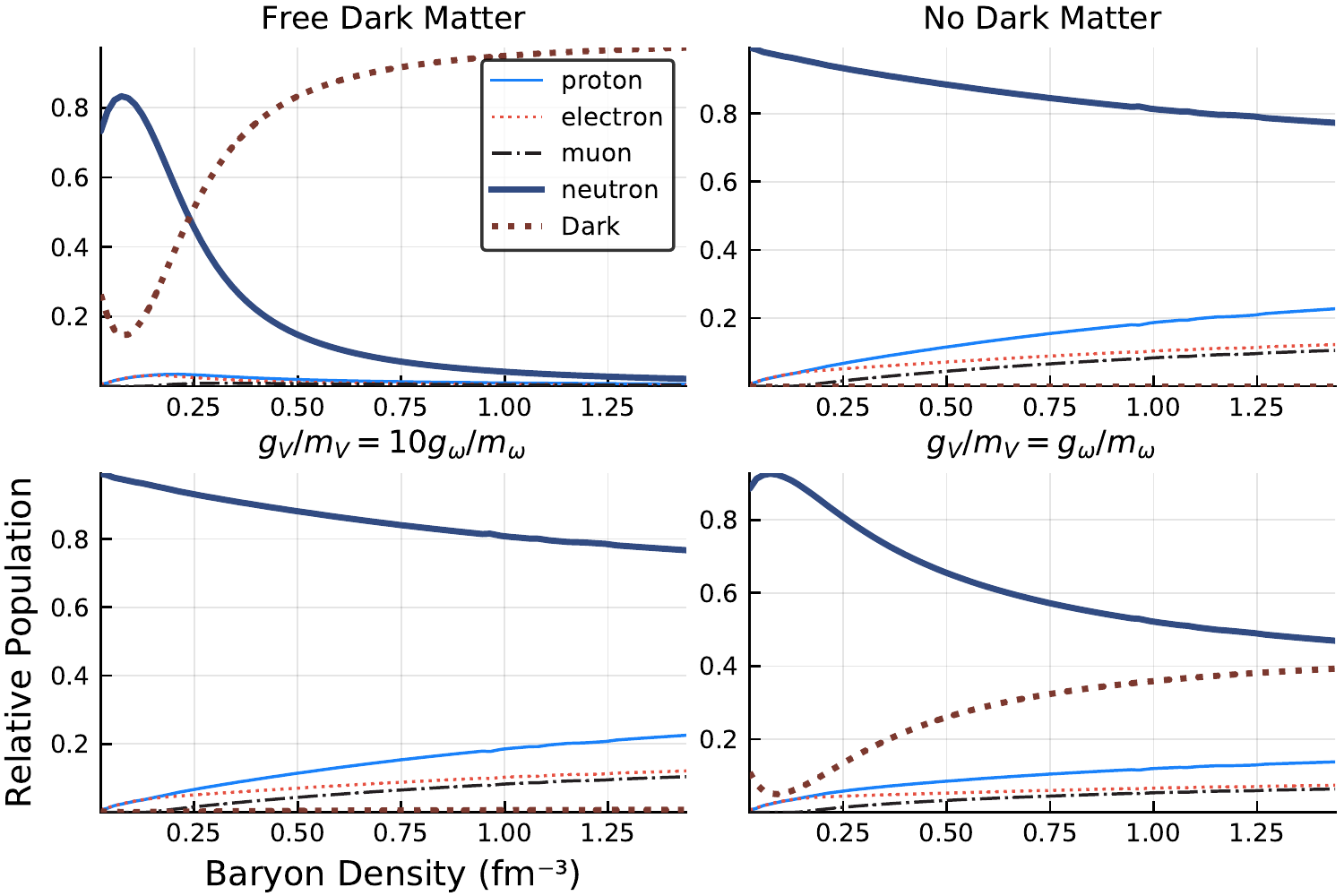}
		\caption{Species fraction considering different strengths of vector self-repulsion.}
		\label{fig:vectorfraction}
	\end{center}
\end{figure}

\begin{figure}[ht]
	\begin{center}
		\includegraphics[width=4in]{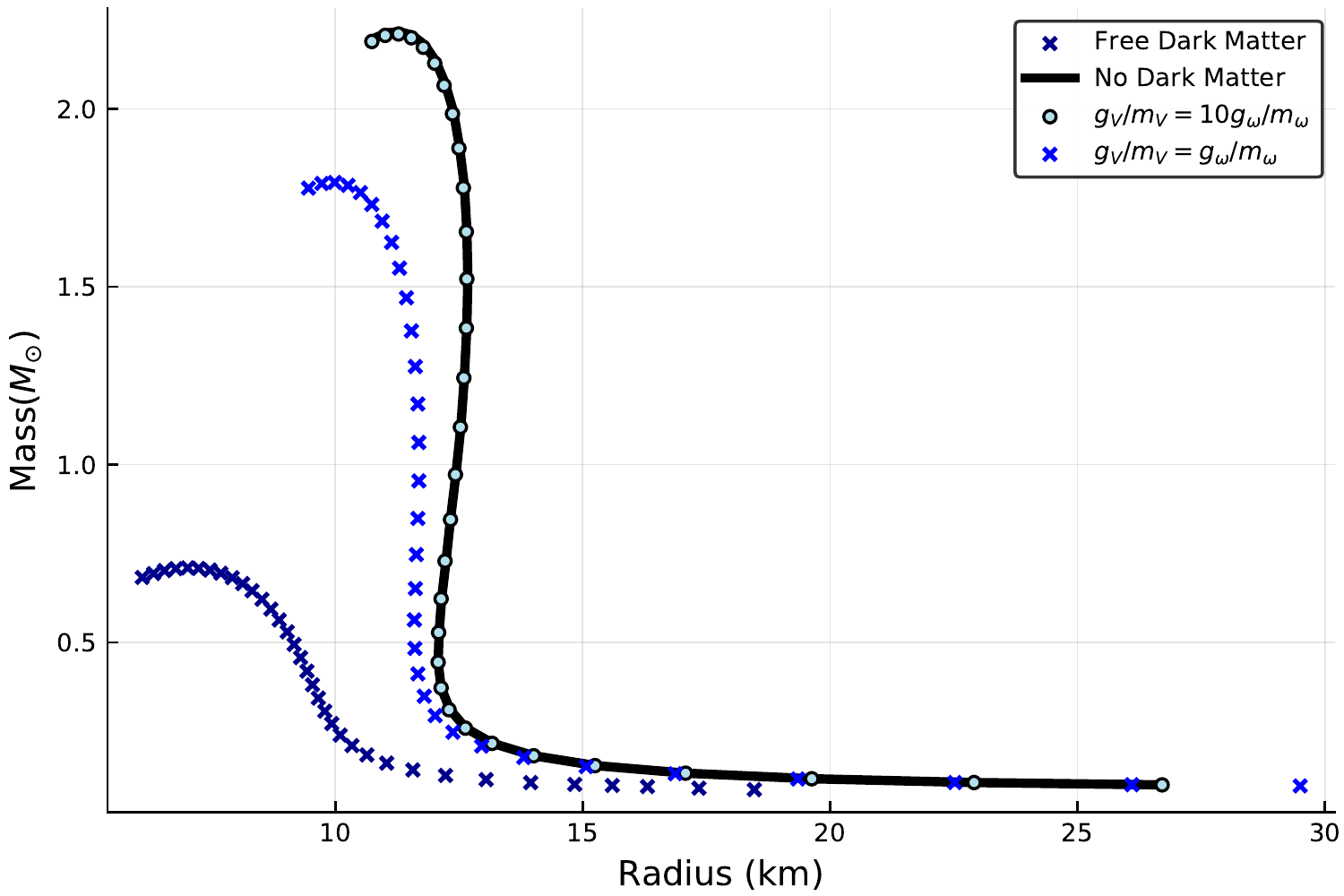}
		\caption{Adding vector self-interactions between the dark fermions through the exchange of a vector boson.}
		\label{fig:vectorinteraction}
	\end{center}
\end{figure}

The species fraction changes as the $\chi V$ interaction becomes stronger and therefore restores the EOS to it's previous stiffer version. The greater the strength of the interaction, the less dark matter will be present in the star (Fig.\ref{fig:vectorfraction}). That allows the system to support much higher masses.
The maximum mass gets to the 2 solar masses value, as all neutron star models must per recent experimental determinations\cite{Antoniadis,Demorest}, only when the $g_V/m_V$ for the dark matter is 10 times greater than $g_\omega/m_\omega$ (Fig.\ref{fig:vectorinteraction}).
However, one must consider that Ref.~\cite{DAmico} severely limits the cross-section of such a dark matter particle through astrophysical data recently measured~\cite{Barbecue}. These values of couplings (that is $g_V/m_V$) are way to high to even enter consideration.

\section{Conclusion}
We have shown that the addition of this dark matter particle to the composition of neutron stars leads to a giant decrease in maximum mass. The mass versus radius diagram points to $0.7$M$_\odot$ as the mass upper limit for stars with dark matter, however further investigations suggest that, if a star with dark matter is a decay product of a normal neutron star the real maximum mass has to be around $0.58$M$_\odot$. Since most neutron stars measured have masses around $1.5$M$_\odot$ this points to a clear inconsistency of the hypothesis with data. Moreover, a repulsive self-interaction indeed can push the mass limit to an acceptable point only when the ratio coupling/mass of the $\chi V$ interaction is 10 times larger than the $n\omega$ vertex. If this dark matter particle were to correspond with astrophysical dark matter this result would be in clear contradiction to recent astrophysical measurements, as pointed out in Ref.~\cite{DAmico}. Even if it were unconnected with astrophysical dark matter, it would be truly remarkable to have a new kind of matter with self interactions an order of magnitude larger than the familiar strong force.

We therefore state that this decay is simply in contradiction with the data of neutron star masses if the dark particle in neutron decay is a significant component of the dark matter in the universe.

\section*{Acknowledgements} 

This work was supported by the University of Adelaide and by the Australian
Research Council through the ARC Centre of Excellence for Particle Physics at
the Terascale (CE110001104) and Discovery Project DP150103164.

\bibliographystyle{ws-procs961x669}
\bibliography{biblio}
\end{document}